\documentclass{amsart}

\newcommand{\LT}[1]{\mbox{{\rm\small LT}$(#1)$}}
\newcommand{\mfrac}[2]
{\raisebox{0.1em}{\mbox{\footnotesize$\displaystyle
\frac{\raisebox{-0.1em}{\mbox{$#1$}}}{#2}$}}}

\newcommand{\wpA}{{\wp}_{\!\!{}_A}}
\newcommand{\wpO}{{\wp\phantom{'}}_{\!\!\!{}_\Omega}}
\newcommand{\wppO}{{\wp'}_{\!\!\!{}_\Omega}}

\newcommand{\wpalpha}{\wp_{\!\raisebox{-0.3ex}{\mbox{\scriptsize$\alpha$}}}}
\newcommand{\wppA}{{\wp'}_{\!\!\!\!{}_A}}
\newcommand{\wppalpha}{{\wp'}_{\!\!\!\raisebox{-0.1ex}
                                      {\mbox{\scriptsize$\alpha$}}}}

\begin{document}

\title{Elliptic Solitons and Gr\"obner Bases}

\author{Yurii V.\,Brezhnev}
\address{Theoretical Physics Department,
Kaliningrad State University,
A.\,Nevsky st.\;14, Kaliningrad 236041,  Russia.}
\email{brezhnev@mail.ru}
\curraddr{Department of Mathematics, Heriot--Watt University,
Edinburgh EH14 4AS, United Kingdom.}
\email{brezhnev@ma.hw.ac.uk}


\date{1 April 2002}


\begin{abstract}
We consider
the solution of spectral problems with elliptic coefficients in the
framework of the Hermite ansatz.  We show that the search for
 exactly solvable potentials and their spectral characteristics 
is reduced to a system
of po\-ly\-no\-mial equations solvable by the Gr\"obner bases method and
others.  New integrable potentials and corresponding solutions of the
Sawada--Kotera, Kaup--Kupershmidt, Boussinesq equations and others
are found.
\end{abstract}

\maketitle
\section{Introduction}
The paper is devoted to the algorithmic problems
associated with integrating  the spectral problems
\begin{equation}
\widehat L\Psi\equiv\frac{d^n}{dx^n}\Psi(x;\lambda)+u_1^{}(x,\lambda)\,
\frac{d^{n-1}}{dx^{n-1}}\Psi(x;\lambda)
+\cdots+u_n(x,\lambda)\,\Psi(x;\lambda)=0,
\end{equation}
where $u_j^{}$ are elliptic functions of $x$ and arbitrary (rational or transcendental) functions of
$\lambda$.  We shall restrict our consideration to the Schr\"odinger
equation
\begin{equation}
\Psi'' - u(x)\,\Psi=\lambda\,\Psi,
\end{equation}
the equation
\begin{equation}
\Psi''' - u(x)\,\Psi'=\lambda\,\Psi,
\end{equation}
and the generalization of the Halphen equation
\begin{equation}\label{halph}
\Psi''' - u(x)\,\Psi' -v(x)\,\Psi=
\lambda\,\Psi.
\end{equation}
We use the term {\em potential\/} for the $u(x),\,v(x)$-functions.  Until
the 1970/80's, few exactly solvable potentials were known.  Earlier,
in 1872, Hermite \cite{hermit} developed an approach for the
integration of the Lam\'e equation
\begin{equation}\label{lame}
\Psi''-n(n+1)\,\wp(x)\,\Psi=\lambda\,\Psi,
\end{equation}
and later, Halphen extended it to the third order equation
\begin{equation}\label{halphen}
\Psi'''-(n^2-1)\,\wp(x)\,\Psi'-\mfrac12\,(n^2-1)\,\wp'(x)\,\Psi=\lambda\,\Psi.
\end{equation}
Here and below $\sigma,\,\zeta,\,\wp,\,\wp'\!$ denote the standard
Weierstrassian functions.  See \cite{halphen, forsyth, kamke} for an
extensive discussion of these classical examples.
According to
modern terminology, the set of exactly solvable elliptic potentials is a
particular case of finite-gap potentials in elliptic functions.  
An intense investigation of
elliptic potentials was initiated by the paper \cite{krichever}, and
in 1987  Verdier and Treibich \cite{treb1,treb2} unexpectedly found new
potentials for the equation (2) in elliptic functions 
\begin{equation}\label{tr}
u(x)=6\,\wp(x)+2\,\wp(x-\omega_j^{})
\end{equation}
and gave the term {\em elliptic solitons} to them.
Recently V.\,Matveev drew attention to the fact that such potentials,
in Jacobian form, were already considered by Darboux in a short note 
\cite{darboux} in 1882.  The following year, two comprehensive m\'emoires
by Sparre  \cite{s1} appeared on this topic.
  
The development of a theory led to the
current result that elliptic solitons are the widest class of 
finite-gap {\em explicit\/} solutions. See, for example, \cite{smirnov1,
enol1}, recent results in \cite{smirnov2,treb3} and references
therein.  The monograph \cite{monograph} reviews work in finite-gap
theory up to the beginning of the 1990's and the review \cite{gesztesy}
and preparing book \cite{gesztesy3} 
contain a wide bibliography on that score.

One feature of elliptic solitons is the potential, and the
$\Psi$-function can be found by the Hermite ansatz method
\cite{hermit}.  In the case of the potentials with the only pole in a parallelogram
of periods, the derivation
of the algebraic curve and other characteristics is not difficult.
For this purpose it is enough to take a few resultants \cite{enol2,
monograph}, but in the general case the elimination technique is
insufficient.  Sect.\,2 contains a pure algebraic interpretation of
Hermite's method.  In Sect.\,3, we show that the general scheme for
solving the problem under consideration (1) is reduced to the 
computation of the Gr\"obner
basis for some polynomial system.  After Buchberger's discovery in
1960's of an algorithm for finding the polynomial ideal bases, this
area of algorithmic mathematics  has rapidly
developed.  See \cite{progr} with regard to the modern achievements in
this area.  Sect.\,4 contains a relation between the method and traditional
objects in finite-gap integration theory: the canonical form of an 
algebraic curve
$\widetilde F(\mu,\lambda)=0$ and reduction of one of the holomorphic 
differentials
to the elliptic. Some new examples of elliptic solitons and their
applications to the  integrable partial differential equations
(\textsc{pde}'s) are presented in Sects.\,5--6 and development of the theory 
is discussed in Sect.\,7.

\section{Algebraic characterization of the Hermite method}
Based on the  $\Phi$-function
\begin{equation}
\label{phi}
\Phi(x;\alpha)=\frac{\sigma(\alpha-x)}{\sigma(\alpha)\,\sigma(x)}\,
e^{\zeta(\alpha)\,x}
\end{equation}
or more precisely {\em l'\'el\'ement simple}
$$
\mathbf\Phi(x;\alpha, k)=\frac{\sigma(\alpha-x)}{\sigma(\alpha)\,\sigma(x)}\,
e^{(\zeta(\alpha)+k)\,x}
$$
by Halphen \cite{halphen1, halphen2},
Hermite and Halphen \cite{hermit, halphen} considered the following ansatz
for the $\Psi$-function
\begin{equation}
\label{anz}
\Psi=\mathbf\Phi(x;\alpha,k)+a_1^{}\mathbf\Phi'(x;\alpha,k)+\cdots+
a_n^{}\mathbf\Phi^{(n)}(x;\alpha,k).
\end{equation}
The function $\mathbf\Phi(x;\alpha,k)$ as well its $x$-derivatives 
$\mathbf\Phi'(x;\alpha,k),\,\ldots,\,\mathbf\Phi^{(n)}(x;\alpha,k)$ are
doubly-periodic functions of $x$ of the second kind.  According to (1) and
(\ref{phi}--\ref{anz}), the expression 
\hbox{$\widehat L\Psi\!/\mathbf\Phi(x-x_0^{};\alpha,k)$} is a 2-periodic
meromorphic function with  only one simple pole at the point
$x_0^{}$. It must be a constant function.  Setting it to be equal to zero,
we  have $\widehat L\Psi=0$ under the corresponding choice of additional
parameters
$k,\,\alpha$ and $a_j^{}$. As the $\mathbf\Phi(x;\alpha,k)$-function has the first
order pole at $x=0$ \cite[p.\,231]{halphen1}
$$
\mathbf\Phi(x;\alpha,k)=\frac1x+k+\frac{k^2-\wp(\alpha)}{2}\,x+
\frac{k^3-3\,\wp(\alpha)\,k+\wp'(\alpha)}{6}\,x^2+
\cdots
$$
to  solve the problem, it is sufficient to equate to
zero only the principal part(s) of the Laurent's expansion(s) of the expression 
$\widehat L\Psi$, where $\Psi$ is the ansatz (\ref{anz}) or its multi-pole 
generalizations (see {\bf Examples 1--3}, {\bf 7} in Sect.\,5).

As a nontrivial example we shall consider the 5-gap Lam\'e potential
with $n=5$ in (\ref{lame}). It has been studied in \cite{enol2}, but we
give a more simple solution.
Note that the cases $n=2,3,4$ and partially 5 were considered already in
\cite[pp.\,527--531]{halphen2}. In the same place one can find mention of
eliminations.  

By Frobenius theory, $\Psi$ must  have a 5-th order pole at $x=0$ and therefore 
the ansatz for the $\Psi$-function should be the following:
\begin{equation}\label{psi}
\Psi=\mathbf\Phi+a_1^{}\mathbf\Phi'+\cdots+
a_4^{}\mathbf\Phi^{\mbox{{\sc\tiny (iv)}}}.
\end{equation}
Substituting (\ref{psi}) in (\ref{lame}) and expanding the result at
$x=0$, we obtain a system of equations in the variables
$k,\,\alpha,\,a_j^{}$. This system is linear  with respect to
$a_j$. We do not write  expressions for the $a_j^{}$ (\cite[formula
(3.7)]{enol2}).  The remaining equations have the form
{\small
$$
\!\!\!\begin{array}{l}
w_1^{}\equiv-6\,k^5+\mfrac{20}{3}\,(9\,\wp+\lambda)\, k^3-60\,\wp'\,k^2+
\left(90\,\wp^2-20\,\lambda\,\wp-\mfrac{10}{7}\,\lambda^2+\mfrac{144}{7}\,
g_2^{}\right) k-{}\\
 \qquad\;{}-\mfrac{4^{\displaystyle{\mathstrut}}}{3} \left(9\,\wp-5\,\lambda\right)\wp',\\\\
w_2^{}\equiv-5\,k^6+\left(75\,\wp+5\,\lambda\right)k^4-100\,\wp'\,k^3+
\left(225\,\wp^2-30\,\lambda\,\wp-\mfrac{5}{7}\,\lambda^2+\mfrac{180}{7}\,
g_2^{}\right)k^2-{} \label{w12} \\
\qquad\;{}-20\left(3\,\wp-\lambda\right)\wp'\,k+25\,\wp^3
-15\,\lambda\,\wp^2+\mfrac{5^{\displaystyle{\mathstrut}}}{7} (\lambda^2-20\,g_2^{})\,
\wp-40\,g_3^{}-\mfrac{1}{21}\,\lambda^3+\mfrac{44}{7}\, 
g_2^{}\,\lambda,
\end{array}
$$}%
which are understood to be equal to zero.  The argument $\alpha$ in
the $\wp,\,\wp'$-functions is omitted for brevity.  The system
(\ref{w12}) has to be considered as  algebraic with respect
to $k$ and {\em transcendental\/} in $\alpha$. We emphasize that everywhere in the paper 
$\lambda$ is a parameter, but not variable in polynomial bases.
Insomuch as functions $\wp(\alpha)$ and $\wp'(\alpha)$ are related by the
Weierstrass equation (torus)
\begin{equation}
\label{w3}
w_3^{}\equiv\wp'(\alpha)^2-4\,\wp(\alpha)^3+g_2^{}\,\wp(\alpha)+g_3^{},
\end{equation}
we supplement (\ref{w12}) by (\ref{w3}) and consider
(\ref{w12}--\ref{w3}) as a {\em polynomial\/} system with respect to
independent variables $(k,\,\wp,\,\wp')$. The simplest method of
solution consists of the  elimination of the variable $\wp$
followed by $\wp'$. As a result, we find that $k$ must be a root of the
polynomial
$$
k^4\,(5103\,k^4-945\,\lambda \,k^2+40\,\lambda^2+54\,g_2^{})^4\,
(225\,(27\,g_2^{}-\lambda^2)\,
P_6^2(\lambda)\,k^2+P_9^{}(\lambda)\,P_3^2(\lambda)),
$$
where $P_{6,9,3}^{}(\lambda)$ are some polynomials in $\lambda$ with
degrees 6, 9, 3 respectively.  It is not difficult to guess that the
correct result requires  that $k$ and $\lambda$ are related
by the following equation
\begin{equation}\label{curve}
F(k,\lambda)\!: \quad 225\,(27\,g_2^{}-\lambda^2)\,
P_6^2(\lambda)\,k^2+P_9^{}(\lambda)\,P_3^2(\lambda)=0,
\end{equation}
since the differential equation (\ref{lame}) is of second order and we must 
have not more than 2
solutions for $k$ with fixed $\lambda$.  Curve (\ref{curve}) can be
brought into the canonical hyperelliptic form
$$
\widetilde F(\mu, \lambda)\!: \quad \mu^2=(27\,g_2^{}-\lambda^2)\,P_9^{}(\lambda)
$$
by an obvious birational transformation (note a misprint $27\,g_2^2$ in this
formula in \cite[formula (3.8)]{enol2}). The variables $\wp,\,\wp'$ as
functions of $k$ can be found in the same way: by the sequential
reduction of exponents of $\wp,\,\wp'$ in (\ref{w12}--\ref{w3}).

Obviously, the resultant technique is almost impossible if the potential
has several poles \cite{enol1,smirnov1}, as the number of variables
increases.   Another approach consists of finding an equivalent
system with the following criterion.  
It is advisable for the new system to contain  linear equations
in $\wp,\,\wp'$.  These equations define $\alpha$ as a function of
$(k,\,\lambda)$:
\begin{equation}\label{cover}
\wp(\alpha)=R_1^{}(k;\,\lambda), \qquad \wp'(\alpha)=R_2^{}(k;\,\lambda)
\end{equation}
and we call (\ref{cover}) a cover of torus (\ref{w3}) in
{\em algebraic\/} form.  Suppose one of the new equations does not
contain $\wp,\,\wp'$ (i.e. be an univariate polynomial in $k$) 
if the nontrivial solution for $k$ exists. We
interpret such a polynomial as {\em the algebraic curve\/} $F(k,\lambda)=0$
corresponding to an elliptic potential. 
If $F(k,\lambda)$ has a factorised form then each of factors is investigated
separately. The curve is one of them. It is clear that its 
degree in $k$ has to  be equal to the order $n$ of the equation (1).
The canonical form 
$\widetilde F(\mu, \lambda)=0$ 
of the curve is obtained with the help of a birational transformation
between variables $(k,\lambda) \leftrightarrow (\mu,\lambda)$ 
(see an explanation in Sect.\,4).
Note, there are specialized algorithms for the computation
of the univariate polynomial in an ideal without solving the system 
as a whole. 

\section{Gr\"obner bases, curves and covers}
We clarify the main idea using the previous example.  Let us consider
three polynomials $w_{1,2,3}^{}(\wp'\!,\wp,k)$
as a system generating an ideal in a polynomial ring 
$\mathbb{Q}(\lambda,g_2^{},g_3^{})[\wp'\!,\wp,k]$
$$
\langle I\rangle=h_1^{}\,w_1^{}+h_2^{}\,w_2^{}+h_3^{}\,w_3^{},
$$
where $h_j^{}=h_j^{}(\wp'\!,\wp,k)$ are arbitrary elements of the ring. 
As is well known, the
structure of the solution of the polynomial systems  depends on
the monomial ordering in a ring \cite{progr}.  The
arguments at the end of Sect.\,2 (see also the elimination theorem in 
\cite{cox}) lead to the choice of pure
lexicographic ordering
$ \wp'\succ\wp\succ k$.
The monograph \cite{cox} contains a good exposition of details of this subject.
The main property of the Gr\"obner base is expressed in

\noindent
{\bf Definition \cite{cox}.} 
\textit{Let $\{w_1^{},\,w_2^{},\,\ldots\}$ be a basis of ideal
$ I=\langle w_1^{},\,w_2^{},\,\ldots\rangle $.
Let $\,\succ\,$ be a monomial ordering on the ring $\mathbb{Q}[\ldots]$ and
\,\LT{f}  denote the leading term $($monomial$\,)$ of a polynomial 
$f \in \mathbb{Q}[\ldots]$.
The set $G=\{f_1^{},\, f_2^{},\,\ldots,\,f_N^{}\}$ is said 
to be a standard
basis $(\mbox{Gr\"obner Basis}\,)$ if the monomial ideal generated by
$$
\big\langle \LT{f_1^{}},\,\LT{f_2^{}},\,\ldots,\,\LT{f_N^{}}\big\rangle
$$
is coincident with an ideal $\langle \LT{I} \rangle$ generated by 
all the leading terms of $I$}.

\noindent
In other words, the leading term of any polynomial in $I$ is divisible
by one of the \LT{f_j^{}}.  According to the definition at the end of Sect.\,2,
the polynomial $F(k;\lambda)$, determining the algebraic curve
\begin{equation}\label{cur2}
F(k;\lambda)=0,
\end{equation}
is a generator of the intersection of the ideal $I$
and the  ring of all polynomials in $k$:
$$
\big\langle F(k;\lambda)\big\rangle=I \,
\raisebox{-0.2em}{\mbox{\LARGE $\cap$}}\,  \mathbb{Q}(\lambda)[k].
$$
Thus we arrive at a general  recipe for the solution of the spectral problem (1).

\noindent
{\bf Proposition.} \textit{Let $\{w_1^{},\,w_2^{},\,\ldots\}$ be polynomials in 
$\wp'(\alpha),\,\wp(\alpha),\,k,\,\ldots$ 
appearing in the Hermite method and 
determining the solution of the spectral problem $(1)\!:$ the cover of
torus {\rm (\ref{cover})} and the curve $(\ref{cur2})$. Then
\begin{enumerate}
\item[$1)$]
Algorithmically, the method of solution is reduced to the computation
of a standard basis for the ideal $I=\langle w_1^{},\,w_2^{},\,\ldots
\rangle$ with respect to pure lexicographic ordering
$\wp'(\alpha) \succ \wp(\alpha) \succ k \succ \ldots.$
 $($For example, by  Buchberger's syzygy polynomials
algorithm $\mbox{\rm \cite{buch})};$
\item[$2)$]
Let
$G=\{f_1,\,f_2,\,\ldots,\,f_N\}$ be this basis.  
The algebraic curve $(\ref{cur2})$ and its
 projection  on the torus $(\ref{w3})$ in algebraic
form $(\ref{cover})$ are contained in $G$ if the univariate polynomial
in $k$ and polynomials 
$(\ref{cover})$ exist\/$;$
\item[$3)$]
If $G$ contains a polynomial free from variables 
$k,\,\wp'(\alpha),\,\wp(\alpha)$, then the spectral problem $(1)$ is 
{\textbf{not integrable}} in the framework of Hermite's ansatz.
\end{enumerate}}

\noindent
{\em Proof\/}.

1) The standard basis always exists and  Buchberger's algorithm  terminates
\cite{cox}.

2) Taking the resultants 
of $w_1^{},\,w_2^{},\,\ldots$
we eliminate
variables $\wp,\,\wp'$ and get
polynomial(s) $R(k)$.
It is obvious that $R(k)\in I$. Using the divisibility
$$
\big\langle \LT{f_1^{}},\,\LT{f_2^{}},\,\ldots,\,\LT{f_N^{}}\big\rangle=
\big\langle \LT{I} \big\rangle
$$
and lexicographic ordering, the  equality $R(k)=h(k)\,\widehat f$ has to occur
for some $\widehat f\in G$ and $h(k)\in \mathbb{Q}(\lambda)[k]$
(possibly equal to 1).  Therefore
there exists a polynomial $\widehat f$ depending only on $k$.  Designating 
$\widehat f(k) \equiv F(k;\lambda)$, 
we obtain the curve (\ref{cur2}). If $F(k; \lambda)$ has
a factorized form, then the algebraic curve is one of its factors. Analogously,
if the polynomials (\ref{cover}) exist, then they necessarily 
belong to $G$. 
In the same way, an important formula --- the curve as a cover of the torus 
(\ref{w3}) in  a {\em transcendental\/} form (an equation in $\alpha$)
\begin{equation}\label{transcov}
R\big(\wp'(\alpha), \,\wp(\alpha); \,\lambda\big)=0
\end{equation}
necessarily must be contained in $G$ computed with the ordering 
$k \succ (\wp'(\alpha) \succ \wp(\alpha) \succ \ldots)$, where
permutations inside the brackets is allowed.
Note,  the order of elliptic function (\ref{transcov}) in
$\alpha$  is equal to the order $n$ of the equation (1).
 
3) An existence of   such a polynomial implies a restriction on
the spectral parameter $\lambda$ (see a demonstrative {\bf Example 5}).
\rule{0.51em}{0.51em} 

Note a direct link of the point 3) to a treatment of finite-gap potentials
as Picard's potentials \cite{gesztesy1}.

There are numerous algorithmic methods to solve this
problem. Among them: the Gr\"obner basis method \cite{buch,coh}, the method of
characteristic sets, and an  effective method of elimination
based on the Seidenberg theory \cite{wang1}.  We do not discuss all the
modern achievements in this area. See \cite{progr} and references
therein for details.  Note that the reduction 
of the holomorphic differential $d \alpha$ to the
elliptic one is derived from (\ref{cover}) by the formula
\begin{equation}\label{holo}
d\alpha=\frac{d\wp(\alpha)}{\wp'(\alpha)}=
\frac{F_k^{}\,{R_1^{}}_\lambda-
F_\lambda^{} \, {R_1^{}}_k}{R_2^{}\, F_k^{}}\,d\lambda,
\end{equation}
where subscripts $k,\,\lambda$ denote the derivatives with respect to $k$ and $\lambda$.

\section{Canonical form of curves and holomorphic differential}

Formulas (\ref{cur2}--\ref{holo}) give a noncanonical form of the curve and holomorphic differential,
i.e. expressions in the variables $(k,\,\lambda)$. The canonical variables
$(\mu,\,\lambda)$ in the algebraic curve $\widetilde F(\mu,\lambda)=0$ we
call variables $\lambda$  in (1) and eigenvalue $\mu$ of a commuting operator pencil
\begin{equation}\label{com}
\widehat P(\lambda)\,\Psi=\mu\,\Psi.
\end{equation}
Supplementing the polar expansion of the equation (1) by the polar expansion
of (\ref{com}), we get the algebraic equations in variables
$(\wp,\,\wp',\,k,\,\mu,\,\ldots)$. Again, based on the above properties of the
Gr\"obner base, a canonical representation of the solution and all the spectral
characteristics  are extracted by computing  the base with the ordering
$(\wp \succ \wp' \succ\ldots) \succ k \succ \mu$. Such a base  contains
a birational transformation between the $(k,\,\mu)$-variables in one direction:
\begin{equation}\label{coor_k}
\mu \to k:\quad k=R_3^{}(\mu;\lambda).
\end{equation}
An inverse transformation 
\begin{equation}\label{coor_mu}
k \to \mu:\quad \mu=R_4^{}(k;\lambda),
\end{equation}
where $R_{3,4}^{}$ are rational functions of its arguments, is computed
by the ordering $(\wp \succ \wp' \succ\ldots) \succ \mu \succ k$.

We add a few words about the efficiency of   computations. The solution of a spectral 
problem itself does not require the  inclusion of a commuting operator (\ref{com}).
So, among of its polar expansions one may take (and supplement) those, 
containing only the $\mu$-variable. Evidently, it will enter into the polar expansion
with first degree: 
\begin{equation}\label{tmp}
\mu = w(k,\wp,\wp';\lambda).
\end{equation}
After the computation  of the base (not including (\ref{com})), 
we will have the curve (\ref{cur2}) and cover in algebraic form (\ref{cover}). Substituting
it into the equation (\ref{tmp}), the pair of equations for the
determination of the above  transformation 
(\ref{coor_k}, \ref{coor_mu}) is
\begin{equation}\label{temp}
\mu = w\big(k,\,R_1^{}(k;\lambda), \,R_2^{}(k;\lambda);\,\lambda\big), 
\qquad F(k;\lambda)=0.
\end{equation}
Formulae (\ref{coor_k}, \ref{coor_mu}) are obtained by  computation
of the bases for (\ref{temp}) with ordering  $(k \succ \mu)$ and $(\mu\succ k)$
respectively. See {\bf Example  6} for  details.

\section{Examples and applications}
In this section we demonstrate the ideology of Sects.\,3--4 on examples.
The generality of the tech\-ni\-que allows us to make further proofs. Let us 
prove that the well known potential of Treibich and Verdier
(\ref{tr}) \cite{treb1} for the equation (2) is the only possible  2-pole potential in the class
\begin{equation}\label{treb}
u(x)=6\,\wp(x)+2\,\wp(x-\Omega),\qquad \Omega\neq 0.
\end{equation}

{\bf Example 1. The Treibich--Verdier potential.}
Parameters $\lambda,\,g_2^{},\,g_3^{}$ are fixed and $\Omega$ is an unknown constant.
The ansatz for the $\Psi$-function must be the following:
\begin{equation}\label{anz1}
\Psi=a_0^{}\mathbf\Phi(x;\alpha,k)+a_1^{}\mathbf\Phi'(x;\alpha,k)+a_2^{}
\mathbf\Phi(x-\Omega;\alpha,k).
\end{equation}
Substituting (\ref{treb}, \ref{anz1}) into (2) and equating the poles
to zero, we obtain $\Psi$-function
$$
\Psi=6\,\mathbf\Phi(-\Omega;\alpha,k)\,\mathbf\Phi'(x;\alpha,k)-
(3\,k^2-3\,\wpalpha-2\,\wpO-\lambda)\,\mathbf\Phi(x-\Omega;\alpha,k)
$$
and a system of five polynomials:
\begin{eqnarray}
w_1^{}&=&2\, (\wpalpha-\wpO )\, k^3+3\,
 (\wppO- \wppalpha )\,{k}^{2}+
2\, (\wpalpha-\wpO ) (3\,\wpalpha-\lambda-2\,\wpO )\,k+\nonumber\\
&&+\,\,(6\,\wpO-\wpalpha)\,\wppalpha-(\wppO-\wppalpha)\,\lambda-(7\,\wpalpha-
2\,\wpO)\,\wppO \nonumber\\
\;\;\;\;\;\;\;w_2^{}&=&3\,k^3-(9\,\wpalpha-4\,\wpO+\lambda)\,k
+3\,\wppalpha+3\,\wppO\nonumber\\
w_3^{}&=&3\,{k}^{4}-2\,(9\,\wpalpha-14\,\wpO-\lambda)\,k^2
+12\, (\wppO+\wppalpha )\,k-\,9\,\wpalpha^{2}-{}\label{w1-5}\\&&
{}-2\, (14\,\wpO+\lambda )\,\wpalpha+12\, 
\wpO^2-8\,\lambda\,\wpO-\lambda^2 \nonumber\\
w_4^{}&=&\wppalpha^2 - 4\,\wpalpha^3 + g_2^{}\,\wpalpha + g_3^{}
\nonumber\\
w_5^{}&=&\wppO^2-4\,\wpO^3+g_2^{}\,\wpO+g_3^{},\nonumber
\end{eqnarray}
where we used the addition theorems for elliptic functions, the important equality
$$
\mathbf\Phi(\Omega;\alpha,k)\,\mathbf\Phi(-\Omega;\alpha,k)=\wpalpha-\wpO,
$$
and designated $\wpalpha \equiv \wp(\alpha),\, \wpO \equiv \wp(\Omega)$, etc.
A common factor $\wpalpha-\wpO$ was
removed in polynomial $w_3^{}$ because it leads to the contradiction:
$\alpha=\Omega$ ($a_2^{}=\infty$).  The system (\ref{w1-5}) generates the
ideal
\begin{equation}\label{ww}
\langle w_1^{},\,w_2^{},\,w_3^{},\,w_4^{},\,w_5^{} \rangle \in 
\mathbb{Q}(\lambda, g_2^{}, g_3^{})[k, \wppalpha, \wpalpha, \wppO, \wpO].
\end{equation}
Note, the same ideal $\langle w_1^{},\ldots, w_5^{} \rangle$ in the ring 
$\mathbb{Q}(\lambda, g_2^{}, g_3^{}, \wpO)[k, \wppalpha, \wpalpha, \wppO]$
leads to the just mentioned condition $\wpalpha-\wpO=0$. 
Therefore $\Omega$ is not arbitrary.
Computing the minimal reduced  Gr\"obner basis for (\ref{ww}) with pure 
lexicographic ordering $\wppalpha \succ \wpalpha \succ k \succ
\wppO \succ \wpO$,
we obtain 8 polynomials, some of them having a factorized form.
If some of the factors do not depend on $\lambda$, we obtain
restrictions on $\wppO$ and $\wpO$ 
equating these factors to zero.  There are four  such polynomials:
$$
\begin{array}{ccl}
G_1 &=& \wppO \left((\lambda^3-4\,g_2^{}\,\lambda-16\,g_3^{})\,k
+(3\,\lambda^2- 4\,g_2^{} )\,\wppO \right) M,\\
G_2 &=& \wppO \left(16\,\wppO\,k+3\,\lambda^2-4\,g_2^{} \right)
^{\displaystyle{}^{\mathstrut}} M,\\
G_3 &=& (4\,\wpO^{3^{\displaystyle{}^{\mathstrut}}}-g_2^{}\,\wpO-
g_3^{})\,(4\,\wpO-\lambda)\,M,\\
G_4 &=& \wppO\,(4\,\wpO-\lambda)^{\displaystyle{}^{\mathstrut}}\,M,
\end{array}
$$
where the multiplier $M$ denotes  $3\,k^2-\lambda-5\,\wpO$.
The equation $M=0$ yields the trivial result $\wp(\alpha)=\wp(\Omega)$.
It is checked by  recomputing the base (\ref{w1-5}) with an additional polynomial $M$.
Further, $\Omega$ must not depend on $\lambda$\,(!).
Therefore, the only solution for $\Omega$ is defined by the equation
$$
\wp'(\Omega)=0 \quad \Longrightarrow\quad 
\Omega=\omega_1^{},\,\omega_2^{},\,\omega_3^{},
$$
where $\omega_j^{}$ are the half-periods of elliptic functions.  Substituting
$\wppO=0,\,\wpO=e_1^{}$, $g_2^{}=4\,(e_1^2+e_1^{} e_2^{}+e_2^2)$, 
$g_3^{}=-4\,e_1^{}e_2^{}(e_1^{}+e_2^{})$ into (\ref{w1-5}) and
recomputing the basis with respect to the ordering
$\wppalpha \succ \wpalpha \succ k$, we obtain the well known
algebraic curve of genus 2 and all algebraic-geometric objects
\cite{belokolos,smirnov1}.  For classification results of the
Treibich--Verdier potentials and other elliptic ones, see
\cite{smirnov1}, the appendix in \cite{JPhysA} and the most recent
results in the review \cite{treb3}.

{\bf Example 2.} As a preliminary,
we shall consider equation (3) with potential 
\begin{equation}\label{new}
u(x)=6\,\wp(x)+6\,\wp(x-\Omega)
\end{equation}
and the restriction
$g_2^{}=0$. As before, we have the ansatz for the $\Psi$-function
$$
\Psi=\mathbf\Phi(x;\alpha,k)+a_1^{}\mathbf\Phi(x-\Omega;\alpha,k)
$$
and the original basis of the ideal is generated by 5 polynomials.
Computing the Gr\"obner basis $G$, we obtain a system of 8
polynomials. Only two of them have a factorized form.
\begin{eqnarray}
G_1&=&
\big(64\,\lambda^3 k^3-27\,(\lambda^2+16\,g_3^{})^2\big)
\big(\wppO\,k-3\,\wpO^2\big ),\nonumber\\
G_2&=&
\left (64\,\lambda^3 k^3-27\, (\lambda^2+16\,g_3^{})^2\right )\!
\left ((4\,\wpO^3-g_3^{}) k-3\,
\wppO\wpO^2\right ).\nonumber
\end{eqnarray}
The nontrivial solution will take place if and only if $(k,\,\lambda)$
be coordinates of the algebraic curve which is the first factor in $G_1,\,G_2$.
In the next example  we rule out the condition $g_2^{}=0$.
 
{\bf Example 3.} If $g_2^{}$ is free, the straightforward computing of
the basis is unsuccessful. Indeed, the Gr\"obner base method is
universal and therefore it can be   ineffective in some special cases. But
our interest is only with the {\em zero structure\/} of the polynomial
system. Thus, the characteristic sets method \cite{wang2} is the best
approach in this case.  Under the ordering
$\wpO \prec \wppO \prec k \prec \wpalpha \prec \wppalpha$,
 the characteristic set has the form
\begin{eqnarray}
f_1^{}&=&\big(64\,(k\,\lambda+g_2^{})(k\,\lambda-2\,g_2^{})^2-27\,
(\lambda^2+16\,g_3^{})^2\big)\,k\,M,\nonumber\\
f_2^{}&=&(8\, (k\lambda-2\,g_2^{} )\,\wpalpha
-8\,{k}^{3}\lambda+16\,g_2^{}\,{k}^{2}+
3\,{\lambda}^{2{\displaystyle{}^{\mathstrut}}}+48\,g_3^{})
\cdot{}\nonumber\\
&&(4\,\wppO k+g_2^{}-12\, \wpO^{2{\displaystyle{}^{\mathstrut}}})\,k,\nonumber\\
f_3^{}&=&\big(32\, (k\lambda-2\,g_2^{} )\,\nonumber
\lambda^{3\displaystyle{}^{\mathstrut}}\, \wppalpha-192\,g_2^2\,k^2\,\lambda^2+{}\\
&&{}+(\lambda^{4{\displaystyle{}^{\mathstrut}}}-288\,g_3^{}\,{\lambda}^{2}-6912\,g_3^2+2^8\,g_2^3 )\,
 k\,\lambda-{}\\
&&{}-g_2^{}\, (11\,{\lambda}^{4{\displaystyle{}^{\mathstrut}}}+864\,g_3^{}\,{\lambda}^{2}+6912\,g_3^2-
2^8\,g_2^3 )\big)\,k\,M, \nonumber\\
f_4^{}&=& \wppO^2-4\, \wpO^3+g_2^{{\displaystyle{}^{\mathstrut}}}\,\wpO+g_3^{},\nonumber
\end{eqnarray}
where
$$
\begin{array}{ll}
M\equiv& 64\,\wppO\,\wpO\,{k}^{2}-4\, (3\,\wppO\,\lambda-16\,g_2^{}\,\wpO-
12\,g_3^{}+96\, \wpO^3 )\,k + {}\\
&{}+3\,(12\, \wpO^2-g_2^{} ) (\lambda+4\,\wppO)^{\displaystyle{}^{\mathstrut}}.
\end{array}
$$
Factorisation shows that the variable $\Omega$ is 
separated in polynomials $f_{1,2,3}^{}$. 
Therefore $f_1^{}$ 
gives an algebraic curve independent of $\Omega$:
\begin{equation}\label{''}
64\,(k\,\lambda+g_2^{})(k\,\lambda-2\,g_2^{})^2=27\,(\lambda^2+16\,g_3^{})^2.
\end{equation}
The polynomials $\{f_2^{},\,f_3^{}\}$ are an algebraic form of the cover
(\ref{cover}).  However, the genus of the curve (\ref{''}) is unity
and we have a cover of a torus by a torus.  Hence, if moduli of both tori are
equal, then there is a
one-to-one correspondence between the global parameter $\alpha$ of
the torus (\ref{w3}) and the global parameter $\tau$ of the torus
(\ref{''}). The next step is to find it. After the birational change
of variables $(k,\lambda)\leftrightarrow (y, x)$:
$$
k=\frac{3\,y^2 + 2\,g_2^{}\,x+3\,g_3^{}}{4\,y\,x} ,\qquad
\lambda=4\,y
$$
we obtain the canonical form of the curve (\ref{''}) as
$y^2=4\,x^3-g_2^{}\,x-g_3^{}$
with an obvious uniformisation and the equality $\alpha=2\,\tau$.
The final solution of the problem (3, \ref{new}) is as follows:
\begin{equation}\label{fullnew}
\begin{array}{c}
\Psi(x;\lambda)=a\,\mathbf\Phi(x;2\,\tau,k)+
\mathbf\Phi(\Omega;2\,\tau,k)\,
\mathbf\Phi(x-\Omega;2\,\tau,k),\qquad \lambda=-4\,\wp'(\tau),\\
a=
\displaystyle
\zeta(2\,\tau+\Omega)- 2\,\zeta(\tau)-\zeta(\Omega)^
{\displaystyle {}^{\mathstrut}},
\qquad 
k=2\,\zeta(\tau)-\zeta(2\,\tau).
\end{array}
\end{equation}
The passage to the limit $\tau \to \omega_j^{}$ in (\ref{fullnew})
leads to the solution under the condition $\lambda=0$: 
$$
\Psi(x;\lambda=0)=C_1\,\Big(\zeta(x)-\zeta(x-\Omega)\Big)+C_2.
$$
An attempt to integrate the more general potential
$$
u(x)=6\,\wp(x-\Omega_1^{})+6\,\wp(x-\Omega_2^{})+A
$$
with a nonzero constant $A$ failed. However, this point has an
explanation in the theory of nonlinear partial differential
equations. Indeed, the spectral problem (3) is associated with the
Sawada--Kotera (SK) equation \cite{sawada}
\begin{equation}\label{sawada}
u_t^{}=u_{\mathit{xxxxx}}^{} - 5\,u\,u_{\mathit{xxx}}^{} 
-5\,u_x^{}\, u_{\mathit{xx}}^{} + 5\,u^2\,u_x^{},
\end{equation}
By assuming that the poles $\Omega_{1,2}^{}$ depend on time $t$, one
obtains an isospectral deformation of this potential. This simple
calculation yields the stationary solution of (\ref{sawada})
$$
u(x,t)=6\,\wp(x-c\,t)+6\,\wp(x-c\,t-\Omega)+A
$$
with the conditions
$$
c+12\,g_2^{}+5\,A^2+ 60\,A\,\wp(\Omega)=0, \qquad A\,\wp'(\Omega)=0.
$$
Therefore $\big(\Omega=\omega_j$ and $A$ is free$\big)$ or 
$\big(A=0$ and $\Omega$ is free$\big)$.  
In the both cases we obtain a restriction on a velocity $c$ of
two cnoidal waves. See an example in \cite{ustinov} 
for the case $A=0$. Recently, Conte and Musette obtained a
similar result \cite[formula (84)]{conte} and revealed a 
remarkable more general  solution in an old paper of Chazy \cite{chazy} in the
context of the Painlev\'e analysis:
\begin{equation}\label{chazy}
u(x,t)=6\,\wp(x-c\,t-\Omega;\,g_2^{},g_3^{})+6\,\wp(x-\,\widetilde c\,t-
\widetilde\Omega; 
\,\widetilde g_2^{}, \widetilde g_3^{}),
\end{equation}
$$
c=3\,g_2^{}-15\,\widetilde g_2^{}, \qquad \widetilde c=3\,\widetilde g_2^{}-15\,g_2^{}.
$$
Strictly speaking, 
Chazy's solution \cite[p.\,380]{chazy} corresponds to the stationary equation 
(\ref{sawada}) 
and therefore to the case $\widetilde g_2^{}=g_2^{}\,(\widetilde c=c)$ in 
(\ref{chazy}). 
One can show that the potential (\ref{chazy}) is  the  stationary solution of 
a linear combination of 
the equation (\ref{sawada}) and higher SK--equation of the 7-th order 
\begin{equation}\label{7}
\begin{array}{l}
\!\!\!\!\!u_t^{}=u_{7x}^{} -{}\\
\,\,\,{}-7\left(
u\,u_{5x}^{}+2\,u_x^{}\,u_{4x}^{}+
3\,u_{\mathit{xx}}^{}u_{\mathit{xxx}}^{}
-2\,u^2\,u_{\mathit{xxx}}^{}-6\,u\,u_x^{}\,u_{\mathit{xx}}^{}
-u_x^3+\mfrac43\,u^3\,u_x^{}\right)^{{}^{\displaystyle{\mathstrut}}}\!.
\end{array}\!\!\!\!\!\!\!\!\!\!\!
\end{equation}
Sect.\,7 contains  additional information for this potential.
We do not enumerate other 1-pole elliptic potentials
$u(x)=A\,\wp(x)+B$ for the equation (3).  For example, one of them is
$u=30\,\wp(x) \pm 3\,\sqrt{3\mathstrut \,g_2^{}}$ (see also \cite{brezhn}).

{\bf Example 4.}
Let us consider a general 1-pole elliptic potential for the equation 
(\ref{halph}) 
\begin{equation}\label{strange}
\Psi'''-( a\,\wp(x)+d)\,\Psi'-
(b\,\wp'(x)+c\,\wp(x) )\,\Psi=\lambda\,\Psi
\end{equation}
in the framework of the ansatz
$$
\Psi=\mathbf\Phi(x;\alpha,k).
$$
Using the above techniques in the ring 
$\mathbb{Q}(a,b,c,d)[\wpalpha,\wppalpha,k]$
we do not get the solution: $I=\langle 1\rangle$. Therefore $(a,b,c,d)$  
have to depend on each
other. After calculations in the ring
$\mathbb{Q}(a,b,c)[\wpalpha,\wppalpha,k,d]$
we determine step by step the  constants $(a,\,b,\,c,\,d)$ and get the following.
The first polynomial in a base is
$$
-8\,b^3\,(2\,b-3)^3\,\lambda^2-
4\,c\,(2\,b-3)^2\,(4\,d\,b^3 -12\,b^2\,d-b\,c^2+6\,c^2)\,\lambda+\cdots=0.
$$
Equating to zero the coefficients in front of $\lambda^2,\,\lambda$ we obtain
$$
b=\frac32, \quad\mbox{or}\quad b=0.
$$
In these cases we will have respectively
$$
3\,d-c^2=0, \qquad (216\,\lambda+c^3-36\,c\,d)\,c^3=0.
$$
Therefore $(b=3/2,\, d=c^2/3)$ or $b=c=0$. In the  first case we have
$$
\Psi'''-3\,(\wp(x)+c^2)\,\Psi'-
\left(\mfrac{3}{2}\,\wp'(x)+3\,c\,\wp(x)\right) \Psi=
\lambda\,\Psi, \quad \quad
\Psi(x;\lambda)=\mathbf\Phi(x;\alpha,c).
$$
The nonramified cover of the torus (\ref{transcov}) of genus $g=1$ is 
$$
\wp'(\alpha)-6\,c\,\wp(\alpha)+2\,\lambda+4\,c^3=0.
$$
$c$ is an arbitrary constant and  the condition
$g_2^{}=0$ \cite[example 3.10]{kamke} does not appear. Note, there is no such 
restriction
in the Halphen equation (\ref{halphen}) with $n=5$ as in \cite{enol2}. 
It appears only for $n=4$ \cite{enol1}. 
The second case is known \cite{kamke}:
\begin{equation}\label{ll}
\Psi'''-(6\,\wp(x)+d)\,\Psi'=\lambda\,\Psi.
\end{equation}
$$
108\,\lambda\,\wp'(\alpha)+36\,(d^2-3\,g_2^{})\,\wp(\alpha)+
27\,\lambda^2-108\,g_3^{}-4\,d^3=0\qquad \mbox{(genus $g=2$)}.
$$
See  \cite{PLA} for an application of this potential.

Note that  both cases can be found in 
\cite[t.\,III: pp.\,372, 522]{hermit} in Jacobian functions 
and \cite[III/IV: pp.\,460, 462]{forsyth} in Weierstrassian functions.
No other possibilities exist. The same technique is applicable to other ansatzs.
The next one is a nontrivial example along these lines.

{\bf Example 5.}
The equation (\ref{strange}) in the framework of the ansatz
\begin{equation}\label{xxx}
\Psi=a_0^{}\mathbf\Phi(x;\alpha,k)+\mathbf\Phi'(x;\alpha,k).
\end{equation}
As a consequence of corresponding indicial equation (\ref{fuchs}) 
with $\nu=2$, without loss of generality we get $b=12-a$ in (\ref{strange}).
Solutions for $a,\,c,\,d$ must not
depend on $\lambda$ and $k$. One solution suggests itself.
Indeed, the first polynomial in the original base has the form
$$
(a-12)\,\big((a-18)\,(k^2-\wpalpha) + 2\,c\,k\big)+ 2\,(a-18)\,d+c^2=0.
$$
With $a=12$, this polynomial does not depend on $k,\,\alpha$ and we get
(after the replacement $c\rightarrow 12\,c$)
$$
a=12,\quad b=0,\quad d=12\,c^2, \quad a_0^{}=k-2\,c.
$$
Moreover, the ideal in the ring 
$\mathbb{Q}(\lambda,c,g_2^{},g_3^{})[\wpalpha,{\wppalpha},k]$ is not equal
to $ \langle 1\rangle$ and therefore, $c$ is an arbitrary constant.
Thus, the equation (\ref{strange}) and its solution take the form
\begin{equation}\label{halphnew}
\begin{array}{c}
\Psi'''-12\,(\wp(x)+c^2)\,\Psi'- 12\,c\,\wp(x)\,\Psi = \lambda\,\Psi,\\
\Psi(x;\lambda)=\mathbf\Phi'(x;\alpha,k)-
2\,c\,\mathbf\Phi(x;\alpha,k)^{\displaystyle{}^{\displaystyle{\mathstrut}}}.
\end{array}
\end{equation}
We do not give here the large formulae for the cover (\ref{cover}),
or the 4-sheet cover in the form (\ref{transcov}) and write only a
skeleton of the non-hyperelliptic trigonal algebraic curve (\ref{cur2}) of genus $g=3$
$$
64\,(\lambda^2+32\,c^3\,\lambda+ 2^8\,c^6- 
108\,g_2^{}\,c^2)(\lambda-11\,c^3)\,k^3
 + (\cdots)\,k^2 + (\cdots)\,k+(\cdots)=0,
$$
where $(\cdots)$ designate some polynomials in
$\lambda,c,\,g_2^{},\,g_3^{}$ with integer coefficients
\cite{brezhn}.  Under $c=0$ we
arrive at the case (\ref{new}) with $\Omega=0$.  

The higher ansatzs (\ref{anz}) are
investigated in a similar manner.  Indeed, by Frobenius theory, if
$\Psi$ has the expansion $\Psi=x^{-\nu}+\cdots$, then $a,\,b$ satisfy
the determining equation
\begin{equation}\label{fuchs}
-\nu(\nu+1)(\nu+2)+a\,\nu+2\,b=0.
\end{equation}
A natural question appears: under what parameters $(a,\,b)$ does the equation
(\ref{fuchs}) have integral solutions for $\nu$\,? 
One of solutions is Halphen's equation (\ref{halphen}). 
It corresponds to $2\,b=a$ and (\ref{fuchs}) is reduced to
$$
a=n^2-1 \quad (n \equiv \nu+1).
$$

As in the previous example we can list all the possible cases for the ansatz (\ref{xxx}). 
Indeed, assuming $b=12-a$ and $(a,d,c)$
to be arbitrary, the origin base contains three polynomials \cite{brezhn}.
The first and second of them are linear in $\wp,\,\wp'$. Solving them and substituting into
the base again, we obtain the remaining polynomial in $(\lambda,\,k)$:
$$
-4\,(a-12)^2\,(a-18)^2\,\big( (a-6)\,(a-18)\,k + c\,(a-9)\big)\,\lambda+
(\cdots)\,k+(\cdots)=0,
$$
where dots denote a polynomial in $(a,\,c,\,d,\,g_2^{},\,g_3^{})$.
It must be zero for all values of $\lambda$.
Splitting it in $\lambda$ we get two linear polynomials in $k$.
Their compatibility condition is the polynomial
$$
(a-6)\,(a-8)\,\big(108\,c^4-72\,(a-18)^2\,d\,c^2+
(a-18)^4\,(12\,d^2-(a-12)^2\,g_2^{})\big)=0
$$
and solution for $k$
$$ 
k=-\frac{(a-9)\,c}{(a-6)\,(a-18)}.
$$
The verifying of Weierstrass's relation (\ref{w3}) yields a polynomial in $\lambda$
$$
(a-6)^3\,(a-12)^3\,(a-18)^6\,\lambda^2+(\cdots)\,\lambda+(\cdots)=0.
$$
Under $a \ne 6,\,12,\,18$ we arrive at the point 3) of the {\bf Proposition}.
Therefore, only three possibilities exist: $a=6,\,12,\,18$.
The corresponding final solutions for the variables $(\wp',\,\wp,\,k)$ are 
obtained  
separately: by recomputing the base. Thus, besides  (\ref{halphnew}),
we have the following
integrable potentials
(note a misprint $\wp(x)$ instead of $\wp'(x)$ in one of the 
formulae in \cite{brezhn})
$$
\Psi'''-(18\,\wp(x)+d)\,\Psi'+6\,\wp'(x)\,\Psi=\lambda\,\Psi,\qquad
\Psi'''-(6\,\wp(x)+d)\,\Psi'-6\,\wp'(x)\,\Psi=\lambda\,\Psi.
$$
See \cite{unterkofler} for  solutions of the generalized
Halphen equation (\ref{halphen})
and \cite{brezhn} for  details of the {\bf Example 5}.

It should perhaps be noted here that the example (\ref{halphnew}) is 
the  generalization $c\ne 0$ of the first nontrivial
case $n=-3$ in a series of  other Halphen's equations 
\cite[p.\,554]{halphen2}
\begin{equation}\label{ttt}
w'''-\mfrac43\,n^2\,w'\,\wp(z)-\mfrac{2}{27}\,n\,(n+3)\,(4\,n-3)\,w\,\wp'(z)=0
\end{equation}
{\em without\/} a spectral parameter\footnote{The example (\ref{ttt}) 
was revealed by E.\,Previato}.
Notation as in \cite[III/IV: {\em Ex\/}.\,15, p.\,464]{forsyth}.
Indicial equation (\ref{fuchs})
for the example (\ref{ttt}) becomes
$$
(3\,\nu+2\,n)(3\,\nu+2\,n+6)(3\,\nu-4\,n+3)=0
$$
and $(n+3)(n+6)(4\,n-9)=0$ for the ansatz (\ref{xxx}) ($\nu=2$).

{\bf Example 6}. Halphen's equation (\ref{halphen}) with $n=5$. 
Here we display only the final formulae in the context of Sect.\,4: 
\begin{itemize}
\item The commuting operator pencil (\ref{com}):
$$
\lambda\,\Psi''-14\,(4\,\wp(x)^2-g_2^{})\,\Psi'+
16\,(7\,\wp'(x)-\lambda)\,\wp(x)\,\Psi=\mu\,\Psi;
$$
\item The polynomial (\ref{tmp}):
{\small
$$
6\,\mu-56\,k^5+560\,\wpalpha\,k^3-20\,(28\,\wppalpha-\lambda)\,k^2+
168\,(5\,\wpalpha^2-g_2^{})\,k\,-4\, (28\,\wppalpha +5\,\lambda)\,
\wpalpha=0;
$$}
\item The birational transformation (\ref{coor_k}, \ref{coor_mu}),
which is quadratic in $(k,\,\mu)$:
$$
\mu=\frac {32}{49}\,\frac {\big(2\,(\lambda^2-392\,g_3^{})\,k\,\lambda+
7\,(5\,\lambda^2-784\,g_3^{})\,g_2^{}\big)\,\big((\lambda^2-392\,g_3^{})\,k-
21\,g_2^{}\,\lambda\big)}{\lambda^{4^{\mathstrut}}-208\,g_3^{}\lambda^2+
3136\,(g_2^3+4\,g_3^2)},
$$
$$
k=\frac {7}{8}\,\frac {\mu^2-4\,g_2^{}\,(5\,\lambda^2-784\,g_3^{})}
{(\lambda^{2^{\mathstrut}}-392\,g_3^{})\,\lambda};
$$
\item The canonical form of the algebraic curve of genus 4 
(see also \cite{unterkofler}):
$$
\widetilde F(\mu,\lambda): \quad
\mu^3-4\,g_2^{}\, (11\,\lambda^2-784\,g_3^{})\,\mu-
\lambda^5 + 208\,g_3^{}\,\lambda^3-3136\,(g_2^3+4\,g_3^2)\,\lambda=0;
$$
\item The 8-sheet cover in the form (\ref{transcov}):
$$
2^8\,(\lambda^2-392\,g_3^{})^3\,\lambda\,\wp'(\alpha)-
2^8 49\,g_2\, (\lambda^2+112\,g_3^{})\,(\lambda^2-392\,g_3^{})^2\,\wp(\alpha)+
\lambda^8-\cdots = 0.
$$
\end{itemize}
Note that  both this cover and its algebraic form (\ref{cover}) are the expansive expressions,
whereas the reduced holomorphic differential (\ref{holo}) in the  variables 
$(k,\,\lambda)$ and $(\mu,\,\lambda)$ is given by the simple formulae:
$$
\frac{d\wp(\alpha)}{\wp'(\alpha)}=
\frac {-8\,(\lambda^2-56\,g_3^{})}
{3\,\mu^2-4\,g_2^{}\,(11\,\lambda^2-784\,g_3^{})} \,d\lambda =
\frac
{-7\,(\lambda^2-56\,g_3^{})}
{(\lambda^2-392\,g_3^{})\,(3\,\lambda\,k+14\,g_2^{})}\,d\lambda.
$$
Analogs of the above formulae are derived for all other examples in the paper.

{\bf Example 7.}
The 2-pole  potential for the equation (\ref{halph}) with ansatz
$$
\Psi=\mathbf\Phi(x;\alpha,k)+
a_1^{}\,\mathbf\Phi(x-\Omega;\alpha,k).
$$
The general 2-pole elliptic potentials contain many parameters ---
multipliers before the $\wp'\!,\, \wp,\,\zeta$-functions. We do not give 
their exhaustive classification and consider only the most interesting 
case
\begin{equation}\label{halph33}
\begin{array}{l}
\Psi'''-3\,\big(\wp(x)+\wp(x-\Omega)-\wp(A)\big)\,\Psi'-{}\\
\;\;\;\quad-\mfrac{3}{2} \Big( \wp'(x)+\wp'(x-\Omega)+
B\,\wp(x)- B\,\wp(x-\Omega)\Big)^{\displaystyle{\mathstrut}}\Psi=\lambda\,\Psi.
\end{array}
\end{equation}
By virtue of the {\bf Proposition}, the Gr\"obner base contains all
the information about the solution, i.e. all the following formulae.
As before we obtain that $\Omega,\,A$ are arbitrary constants and
$$
B=2\,\sqrt{\wp(\Omega)-\wp(A)}.
$$
The parameters $k$ and $\lambda$ as meromorphic functions are related by 
a algebraic
equation of the genus 2 independent of $\Omega$ 
(compare with \cite{smirnov2}):
\begin{equation}\label{curv}
2\,\lambda\, k^3+(3\,\wpA^2-g_2^{})\,k^2-3\,
\wpA\,\lambda \,k-\mfrac14 \,\lambda^2 + \wppA^2=0.
\end{equation}
The equation (\ref{curv}) can be realized as a 2-sheet cover of 
a torus in the form
(\ref{transcov})
$$
\lambda\,\wp'(\alpha)+(3\, \wpA^2-g_2^{})\,\wp(\alpha) 
+\mfrac14\,\lambda^2 +\wpA^3-g_3^{}=0.
$$
The algebraic form of cover (\ref{cover}) has the form
$$
\wp(\alpha)=k^2+\wpA,\qquad
\wp'(\alpha)=
\left(g_2^{}-3\,\wpA^2 \right)\,\frac{k^2}{\lambda}-
\frac{\wppA^2}{\lambda} - \frac{\lambda}{4}.
$$

\section{Solutions of integrable {\sc PDE}'s}
The spectral problem (\ref{halph}, \ref{halph33}) 
corresponds to the Boussinesq equation
\begin{equation}\label{bouss}
3\,u_{tt}^{}=\left(2\,u^2-u_{xx}^{}\right)_{xx},
\end{equation}
and the arbitrariness of $\Omega$ means the existence of an
isospectral deformation of the potential
\begin{equation}\label{deform}
u(x,t)=3\,\wp(x-\Omega_1^{}(t))+3\,\wp(x-\Omega_2^{}(t))-3\,\wpA.
\end{equation}
Substituting the 2-gap ansatz  (\ref{deform}) in (\ref{bouss}),
we get the well known
system of  pairwise-interacting  particles  $\Omega_{1,2}^{}(t)$
of   the Calogero--Moser 
system type \cite{krichever} with a repulsion potential and 
immovable center of mass. 
Integration leads to an equation for $\Omega_1^{}(t)$:
$$
\dot \Omega_1^2=4\,\wp(2\,\Omega_1^{}-c)-4\,\wpA, \qquad 
\Omega_2^{}=c-\Omega_1^{}.
$$
Using the uniformisation of the corresponding elliptic
curve, we obtain the explicit solution
\begin{equation}\label{bus}
\Omega_1^{}(t)=\mfrac12\,\wp^{-1}\!\left(
\widetilde{\zeta}(\tau+\nu)-\widetilde{\zeta}(\tau-\nu)-\widetilde{\zeta}(2\,\nu)+
\mfrac14\,\wpA\right)+\mfrac{c}{2},
\end{equation}
$$
\nu=\pm\,\mfrac12\,\widetilde{\wp}^{-1}\!\left(\mfrac{\wpA^2}{16}+
\mfrac{g_2^{}}{24}\right),
\qquad
\widetilde\wp'(2\,\nu)=-\mfrac{1}{32}\,\big(\wpA^3-g_2^{}\,\wpA +2\,g_3^{}\big),
$$
where $\tau=8\,(t-t_0^{})$ and $c,\,t_0^{}$ are arbitrary constants.  An
implicit form of this solution with $c=0$ in terms of Jacobi's
sn, cn, dn-functions is given in \cite{perelomov} and in the earlier citation
\cite{chud} in the context of solutions of the Kadomtsev--Petviashvili
equation.  In (\ref{bus}), the
elliptic integral $\wp^{-1}$ is calculated with invariants
$g_2^{},\,g_3^{}$, and the $\widetilde{\zeta},\,\widetilde{\wp}^{-1}$-functions with
invariants
$$
\widetilde{g}_2^{}=\frac{1}{16}\left(g_2^{}\, \wpA^2+3\,g_3^{}\,
\wpA + \frac{g_2^2}{12}\right),\qquad 
\widetilde{g}_3^{}=-\frac{g_3^{}}{2^8} \left( 
\wppA^2+2\,g_3^{}\right)+
\frac{g_2^3}{2^7 3^3}-\frac{g_2^{}\,\widetilde{g}_2^{}}{24}.
$$

The reduction case $B=0$ in (\ref{halph33}) corresponds to the
Kaup--Kupershmidt (KK) equation
\begin{equation}\label{kk}
u_t^{}=u_{\mathit{xxxxx}}^{}-5\,u\,u_{\mathit{xxx}}^{}-
\mfrac{25}{2}\,u_x^{}\,u_{\mathit{xx}}^{} +5\,u^2\,u_x^{}
\end{equation}
and its stationary solution
\begin{equation}\label{solkk}
u(x,t)=3\,\wp(x-c\,t)+3\,\wp(x-c\,t-\Omega)-3\,\wp(\Omega),
\end{equation}
but the velocity $c$ depends on the distance between poles:
$$
c=3\,g_2^{}-45\,\wp^2(\Omega).
$$
The generalization of  (\ref{solkk}) in a similar way as the solution 
(\ref{chazy}) is
\begin{equation}\label{last}
u(x,t)=-12\,(\wp_1^{}+\wp_2^{})+3
\left( 
\frac{\wp'_1-\wp'_2}{\wp_1^{}-\wp_2^{}}
\right)^2,
\end{equation}
where
$$
\wp_1^{}\equiv\wp(x-c\,t-\Omega; \,g_2^{}, g_3^{}), \qquad
\wp_2^{}\equiv\wp(x-c\,t-\widehat \Omega; \,g_2^{}, \widehat g_3^{}), \qquad
c=-12\,g_2^{}.
$$
Here $\Omega,\,\widehat \Omega,\,g_2^{},\, g_3^{}, \,\widehat g_3^{}$ are
five arbitrary constants. 
Using a connection between the SK($u$)- and KK($w$)-equations and the Tzitzeica equation
\begin{equation}\label {tz}
\phi_{\mbox{\scriptsize\it xt}}=e^\phi-e^{-2\,\phi}
\end{equation}
with the Fordy--Gibbons equation \cite{fg}
$$
v_t=v_{\mbox{\scriptsize\it xxxxx}}-
5\,\big(v_x\,v_{\mbox{\scriptsize\it xxx}}+
v^2\,v_{\mbox{\scriptsize\it xxx}}+
v_{\mbox{\scriptsize\it xx}}^2+
v_{\mbox{\scriptsize\it x}}^3+
4\,v\,v_x\,v_{\mbox{\scriptsize\it xx}}-
v^4\,v_x \big)
$$
via the Miura transformations
$$
u=v^2-v_x,\qquad w=v^2+2\,v_x, \qquad u=\phi_{\mbox{\scriptsize\it xx}}+
\phi_x^2,
$$
we obtain stationary solutions for the $v$-function
$$
v(x,t)=\frac{{\wp'}_{\!\!1}-{\wp'}_{\!\!2}}{\wp_1-\wp_2}.
$$
Non-stationary solution of the equation (\ref{tz}) has the form
$$
\phi(x,t)=\ln 2\,c+\ln\!\big(\wp(x+c\,t-\Omega; g_2^{}, g_3^{}) - 
\wp(x-c\,t-\widehat\Omega; g_2^{}, \widehat g_3^{})\big)
$$
with the  restriction: $4\,(\widehat g_3-g_3^{})\,c^3=1$.
The  details of calculations (\ref{bus}) are expounded in \cite{brezhn}
and the formula (\ref{last}),  $\Psi$-function for 
the potentials (\ref{chazy}, \ref{last}) in \cite{brezhn,brezhn',ustinov}.

\section{Concluding remarks and discussion}
The investigation of elliptic solitons
can be automated by a  polynomial techniques. 
The Gr\"obner base method provides an unified approach to the solution
of related problems. 
The technique suggested with minor modifications 
is extended to  matrix spectral problems.

As the  {\bf Examples 2--3} and {\bf 5} show, the algebraic curves
can be degenerate  (multiply roots of discriminant).

The general case in {\bf Example 7} for the equation (\ref{halph}) 
\begin{equation}\label{bbb}
\begin{array}{l}
u(x)=a\,\wp(x)+b\,\wp(x-\Omega)+c\,\zeta(x)-c\,\zeta(x-\Omega),\\
v(x)=d\,\wp'(x)+e\,\wp'(x-\Omega)+f\,\wp(x)+g\,\wp(x-\Omega)
+h\,\zeta(x)-h\,\zeta(x-\Omega)^{\displaystyle{}^{\displaystyle\mathstrut}}
\end{array}
\end{equation}
requires additional research.

To all appearances, the example (\ref{halphnew}) 
has to fit into the hierarchy of higher Boussinesq equations,
studied  in full in \cite{dickson1, dickson2}.
Multi-pole potentials are investigated by involving  the addition theorem
for the $\mathbf\Phi$-function:
$$
\mathbf\Phi(x+z; \alpha,k)=\frac12\,\frac{\mathbf\Phi(x; \alpha,k)
\,\mathbf\Phi(z; \alpha,k)}
{\wp_{\!\raisebox{-0.3ex}{\mbox{\scriptsize$x$}}}-
\wp_{\!\raisebox{-0.3ex}{\mbox{\scriptsize$z$}}}}
\left(
\frac {\wppalpha+{\wp'}_{\!\!\!\raisebox{-0.3ex}{\mbox{\scriptsize$x$}}}}
{\wpalpha-\wp_{\!\raisebox{-0.3ex}{\mbox{\scriptsize$x$}}}}-
\frac {\wppalpha+{\wp'}_{\!\!\!\raisebox{-0.3ex}{\mbox{\scriptsize$z$}}}}
{\wpalpha-\wp_{\!\raisebox{-0.3ex}{\mbox{\scriptsize$z$}}}}
\right).
$$

A natural assumption suggests itself: 
{\em all the potentials, obtained by the above method, are  
finite-gap ones.\/} At least, by  construction, all such potentials belong 
to the set of 
exact integrable (explicit $\Psi$ \cite{ustinov}) and the $\Psi$-function  is a 
single-value function on a Riemann surface of the algebraic curve $F(k,\lambda)=0$
and meromorphic function (in $x$) for all values of $\lambda$
(Picard's theorem \cite{gesztesy1,gesztesy2}).
Note that remaining linear independent solutions for the $\Psi$-function
are got by choosing of $k$-branch of  algebraic equation (\ref{cur2}).

If the assumption is valid, then the potentials
for the spectral problems (2--\ref{halph}) are free of residues
(a consequence of $\Theta$-formulas), and therefore
ansatzs for the multi-pole potentials (say (\ref{bbb})) 
do not have to involve the $\zeta$-functions.
This strongly decreases the number of parameters and the computational task.

The potential (\ref{chazy}) 
\begin{equation}\label{'''}
u  =  6\,\wp(x-\Omega; \,g_2^{}, g_3^{}) + 
6\,\wp(x-\widetilde{\Omega} ; \,  \widetilde g_2^{}, \widetilde g_3^{})
\end{equation}
with arbitrary invariants 
$g_2^{},\,g_3^{},\,\widetilde g_2^{}, \,\widetilde g_3^{}$ is a finite-gap one
for the equation (3), but its spectral characteristics can not be obtained
in the framework of elliptic soliton theory.
The corresponding commuting operator pencil is derived with the help of the
equation (\ref{7}) and takes the form
$$
\begin{array}{l}
\big(9\,(u-c_1^{})\,\lambda-3\,u'''+6\,u\,u'+c_1^{}\,u'\big)\,\Psi''-{}\\
{}-\big ( 27\,\lambda^2+9\,u'\,\lambda - u^{{\mbox {\tiny\sc (iv)}}}-
3\,u'{}^2+\mfrac43\,u^{3^{\displaystyle{\mathstrut}}}-c_1^{}\,u''-
c_1^{}\,u^2+27\,c_2^{}\big )\,\Psi'+{}\\
{}+6\,\lambda\,(u''-u^{2^{\displaystyle{\mathstrut}}}+c_1^{}\,u)\,\Psi=\mu\,\Psi.
\end{array}
$$ 
Hence, the canonical form $\widetilde F(\mu,\lambda)=0$ of the associated trigonal curve
of  genus $g=4$ is obtained by elimination of $\Psi$:
{\footnotesize
$$
\!\!\!
\begin{array}{l}
 a^3\, \mu^3 +
3^5 \big( 36\,a^2\,b\,\lambda^4-(a^5-16\,g_2^{}\,a^4+16\,g_2^2\,a^3+
192\,(b-3\,g_3^{})\,a^2\,b+
192\,g_2^{}\,a\,b^2)\,\lambda^2+
{\mathstrut}^{\mathstrut}{\mathstrut}_{\mathstrut}\\
{}+48\,(g_3^{}\,a^5-g_2^{}\,a^4\,b+4\,a^2\,b^3)\big)\, \mu+
3^6\, \big (27\,a^3\,\lambda^7 -216\, (a^3 b-4\,g_3^{}\,a^3+
2\,g_2^{}\,a^2\,b+ 8\,b^3)\,\lambda^5-
{\mathstrut}^{\mathstrut}{\mathstrut}_{\mathstrut}\\
{}-2\, (
a^6+30\,g_2^{}\,a^5-96\,g_2^2\,a^4-8\,(45\,b^2-216\,g_3^{}\,b-
8\,g_2^3+432\,g_3^2)\,a^3+
{\mathstrut}^{\mathstrut}{\mathstrut}_{\mathstrut}\\
{}+576\,(b+6\,g_3^{})\,g_2^{}\,a^2\, b - 2^8 9 \, g_2^2\,a\, b^2-
2^8 3^3  \,(b^4-2\,g_3^{}\,b^3))\,\lambda^3-
{\mathstrut}^{\mathstrut}{\mathstrut}_{\mathstrut}\\
{}-288\, (a^3 - 2\,g_2^{}\,a^2+24\,b^2 )
 (g_3^{}\,a^3- g_2^{}\,a^2\,b+  4\,b^3)\,\lambda \big)=0,
{\mathstrut}^{\mathstrut}{\mathstrut}_{\mathstrut}
\qquad
a \equiv g_2^{}-\widetilde g_2^{}, \quad b \equiv g_3^{}-\widetilde g_3^{}
\end{array}
$$
}%
and the corresponding $\Psi$-function is given by the expression
$$
\Psi(x;\lambda)=\exp\!\int \!\!
\frac 
{\lambda\, F^2  - G\,H + F\,H' - F'\,H}
{G^2 - u\, F^2 - F\,H + F'\,G - F\,G'} \,\mathit{dx},
$$
where prime denotes a derivation in $x$ and
$$
\begin{array}{c}
F  \equiv  -3\,u'''+6\,u\,u'-3\,c_1^{}\,u'+9\,(u-c_1^{})\,\lambda,
\qquad
H  \equiv  6\, (u'' - u^2 + c_1^{}\,u)\,\lambda -\mu,
{}_{\mathstrut}^{\mathstrut}\\
G  \equiv  u^{{\mbox {\tiny\sc (iv)}}}  +  c_1^{}\,(u'' + u^2)
- 3\,u'{}^2 - 9\,u'\,\lambda  - \mfrac{4\mathstrut}{3{\mathstrut}} 
\,u^3 - 27\,(\lambda^2 +c_2^{}),
{}_{\mathstrut}^{\mathstrut}\\
c_1^{}  \equiv  
-12\, \displaystyle \mfrac {g_{3{\mathstrut}}^{\mathstrut}-\widetilde g_3^{}}
{g_2^{}-\widetilde g_2^{}}, 
\qquad
 c_2^{}  \equiv  \displaystyle \mfrac83 \,
\mfrac { ( 3\,\widetilde g_3^{} +  g_{3\mathstrut}^{{}^{\mathstrut}})\, g_2^{} - 
(3\,g_3^{} + \widetilde g_3^{})\,\widetilde g_2^{} }{g_2^{} -  \widetilde g_2^{}}.
{}_{\mathstrut}^{\mathstrut}
\end{array}
$$
One particular case of the potential (\ref{'''}) and more general 
property of finite-gap potentials are discussed in \cite{ustinov}.

The natural generalization of Hermite's method is
to consider nonlinear homogeneous ansatzs for the $\Psi$-function. For instance, 
the quadratic ansatz
$$
\Psi=e^{kx}\,\sum\limits_{j,\,n}\, A_{jn}\,\Phi(x-\Omega_j^{};\,\alpha)\,
\Phi(x-\Omega_n^{};\,\alpha).
$$
However, this does not fit into the framework of  finite-gap integration theory,
because the poles of the potential can depend on the spectral parameter. 
The following example with the transcendental dependence on spectral parameter
elucidates this:
$$
\Psi''=\big(6\,\wp(x)+2\,\wp(x-\lambda)+4\,\wp(\lambda)\big)\,\Psi, \qquad 
\Psi(x;\lambda)=\Phi^2(x;\,\lambda).
$$
Actually, the quadratic (and higher) ansatzs will not give an advantage
due to the relation
$$
\Phi(x;\alpha)^2=-\mathbf\Phi'\big(x; 2\,\alpha,
\zeta(2\,\alpha)-2\,\zeta(\alpha)\big)
$$
and we again arrive at the framework of Hermite's method.

Note that the nonintegrability of equation (1) in context of the point 3) of 
the {\bf Proposition},
nevertheless, can be useful for its integrability with special values of $\lambda$
or for more complex operator pencils with a dependence of 
the potential (say parameters $a,\,b,\,c,\,d$ in {\bf Examples  4--5}) 
on $\lambda$.
The availability of additional constants in the potentials may be considered as a 
family of spectral pencils, and under fixed values of  $\lambda$, 
as new spectral problems. 
For instance, the 2-gap 
Lam\'e potential $u=6\,\wp(x)$ for equation (\ref{lame}) is obtained from example (\ref{ll}) 
with $\lambda=0$ and $d \to \lambda$, $\Psi' \to \Psi$.
A less simple example is to swap the parameters $\lambda \leftrightarrow c$ 
in equation (\ref{halphnew}), 
whereupon one finds the finite-gap operator $\lambda$-pencil
$$
\Psi'''-12\,\big(\wp(x)+\lambda^2\big)\,\Psi'-
(c+12\,\lambda\,\wp(x)\big)\,\Psi=0
$$
with the algebraic curve $F(k,\lambda)=0$ of   genus $g=8$
\cite[formula (15)]{brezhn}.

\section{Acknowledgments}
The author is grateful to Dr. A.\,Smirnov for helpful discussions.  
The equation (\ref{halph33}) was considered
in collaboration with him.  The author thanks 
Prof. R.\,Conte for a copy of an old paper \cite{chazy}.
My special thanks are due to
Profs. E.\,Previato and V.\,Enol'skii for  much attention to the work and 
discussions, and Mark van Hoeij  for discussions of the computer
algorithms. 
Dr.  N.\,Ustinov made some important observations.

The author also is grateful to 
Prof. J.\,C.\,Eilbeck
for  numerous discussions, remarks and hospitality at 
Heriot--Watt University where the paper was improved.

The author was supported by Royal Society/NATO Fellowship
and RFBR (00--01--00782).

\bibliographystyle{amsplain}

\end{document}